\documentclass[aps,twocolumn,superscriptaddress,floatfix,longbibliography]{revtex4-1}
\usepackage{bm,amsmath,amsfonts,amssymb,braket}
\usepackage{latexsym,dsfont,array,layout,mathrsfs,textcase}
\usepackage{graphicx,graphics,subfigure,xcolor}
\usepackage{comment,verbatim}
\usepackage{float}
\usepackage{times}
\usepackage{xspace}
\usepackage{stmaryrd}
\usepackage{multirow}
\usepackage{mathtools}
\usepackage{booktabs}

\usepackage[colorlinks,linkcolor=blue,citecolor=blue,urlcolor=blue]{hyperref}
\begin{document}
\title{Multiple localization transitions and novel quantum phases induced by staggered on-site potential}

\author{Rui Qi}
\affiliation{Beijing National Laboratory for Condensed Matter Physics, Institute of Physics, Chinese Academy of Sciences, Beijing 100190, China}
\affiliation{School of Physical Sciences, University of Chinese Academy of Sciences, Beijing, 100049, China}
\author{Junpeng Cao}
\thanks{junpengcao@iphy.ac.cn}
\affiliation{Beijing National Laboratory for Condensed Matter Physics, Institute of Physics, Chinese Academy of Sciences, Beijing 100190, China}
\affiliation{School of Physical Sciences, University of Chinese Academy of Sciences, Beijing, 100049, China}
\affiliation{Songshan lake Materials Laboratory, Dongguan, Guangdong 523808, China}
\affiliation{Peng Huanwu Center for Fundamental Theory, Xian 710127, China}
\author{Xiang-Ping Jiang}
\thanks{2015iopjxp@gmail.com}
\affiliation{Zhejiang Lab, Hangzhou 311121, China}
\affiliation{Beijing National Laboratory for Condensed Matter Physics, Institute of Physics, Chinese Academy of Sciences, Beijing 100190, China}

\date{\today}

\begin{abstract}
We propose an one-dimensional generalized Aubry-Andr{\'e}-Harper (AAH) model with off-diagonal hopping and staggered on-site potential.
We find that the localization transitions could be multiple reentrant with the increasing of staggered on-site potential. The multiple localization transitions are verified by the quantum static and dynamic measurements such as the inversed or normalized participation ratios, fractal dimension and survival probability.
Based on the finite-size scaling analysis, we also obtain an interesting intermediate phase where the extended, localized and critical states are coexistent
in certain regime of model parameters. These results are quite different from those in the generalized AAH model with off-diagonal hopping, and can help us to find novel quantum phases, new localization phenomena in the disordered systems.

\end{abstract}

\maketitle

%%%%%%%%%%%%%%%%%%%%%%%%%%%%%%%%%%%
\section{Introduction}

The quantum localization has been an important research topic in the condensed matter physics since the pioneer works of Anderson et al. \cite{anderson1958absence,abrahams1979scaling}.
It is argued that the delocalization-localization transition can not happen in low dimension because the
weak disorder can localize the eigenstates \cite{lee1985disordered,evers2008anderson}. However, it is demonstrated that one-dimensional quasiperiodic incommensurate lattices can exhibit the localization transition.
The most famous system is the Aubry-Andr{\'e}-Harper (AAH) model \cite{harper1955single,aubry1980analyticity},
which indeed undergo the localization transition at the critical point due to the existence of self-duality symmetry.

Later, it is found that when self-duality of standard AAH model is broken, there are many variants of standard AAH model, where the localization transition could have an energy-dependent single particle mobility edge,
which separates the extended states from the localized ones due to the breaking of self-duality symmetry \cite{sarma1988mobility,sarma1990localization,biddle2010predicted,ganeshan2015nearest,li2017mobility,luschen2018single,li2020mobility,wang2020one,gonccalves2022hidden,RGM2022,criticalMG2023}.
The existence of mobility edge gives that the system has an intermediate phase where the extended and localized states are coexistence in the energy spectrum.

In the Anderson model, the states after localization transition are always localized with the increasing of the disorder potential.
However, recent studies show that the localization transition in some quasiperiodic systems such as the AAH model with staggered on-site potential
can occur many times \cite{goblot2020emergence,roy2021reentrant,wu2021non,jiang2021mobility,zhai2021cascade,padhan2021emergence,han2022dimerization,wang2023fate,nair2023emergent,vaidya2022reentrant}.
Thus the localization transition can be reentrant.
Some localized states after first localization become extended.
Then the extended states could be localized again by the disorder
and the second localization transition arises.

Recently, the critical states cause much attention \cite{chang1997multifractal,takada2004statistics,liu2015localization,tang2021localization,wang2021many,zhai2020many,li2022observation,shimasaki2022anomalous,zhang2022localization,wang2022mobility,lin2022general,li2023emergent}.
In the AAH model with incommensurate modulations on both the on-site potential and the off-diagonal hopping,
besides the extended states and localized ones, there also exist the critical states which have some wonderful properties
such as certain fractal structures. The complete phase diagram of the system includes the extended phase where all the states are extended,
localized phase where all the states are localized, and critical phase where all the states are critical \cite{jagannathan2021fibonacci}.
Obviously, the system does not have the mobility edge. Another interesting progress is that
the tight-binding model with nearest-neighbor hopping and quasiperiodic on-site potential
has an anomalous mobility edge and a quantum phase coexisting the critical and localized states \cite{liu2022anomalous,lin2022general}.
By proposing a quasiperiodic optical Raman lattice model which includes the hopping, spin-orbital coupling and Zeeman terms,
the coexistent phase of localized, extended and critical states is predicted \cite{wang2022quantum}.

At present, the localization transitions and quantum phases related with AAH model and its generalization have many applications in the
cold-atoms \cite{roati2008anderson,luschen2018single,li2017mobility,an2021interactions}, optical lattices \cite{lahini2009observation,kraus2012topological} and non-Hermitian systems \cite{liu2020non,liu2020generalized,liu2021localization,jiang2021non}. The many-body localization phenomena in the interacting systems are also studied extensively
\cite{iyer2013many,nandkishore2015many,wang2021many,deng2017many,hamazaki2019non,zhai2020many,tang2021localization}.

In this paper, we study a generalized AAH model with off-diagonal hopping and staggered on-site potential.
By using the inversed participation ratio, normalized participation ratio, fractal dimension and the quantum dynamics measurements, we find that the system has multiple localization transitions accompanied with several intermediate phases with the increasing of quasiperiodic potential.
Based on the multifractal analyses of eigenstates and finite-size behavior, we obtain that there indeed exist a quantum phase with coexisting localized, extended, and critical states in certain regime of model parameters. These results are quite different with those in the AAH model with off-diagonal hopping, which only has the localized, extended and critical phases.
\begin{figure*}[htb]
\includegraphics[width=1.0\textwidth]{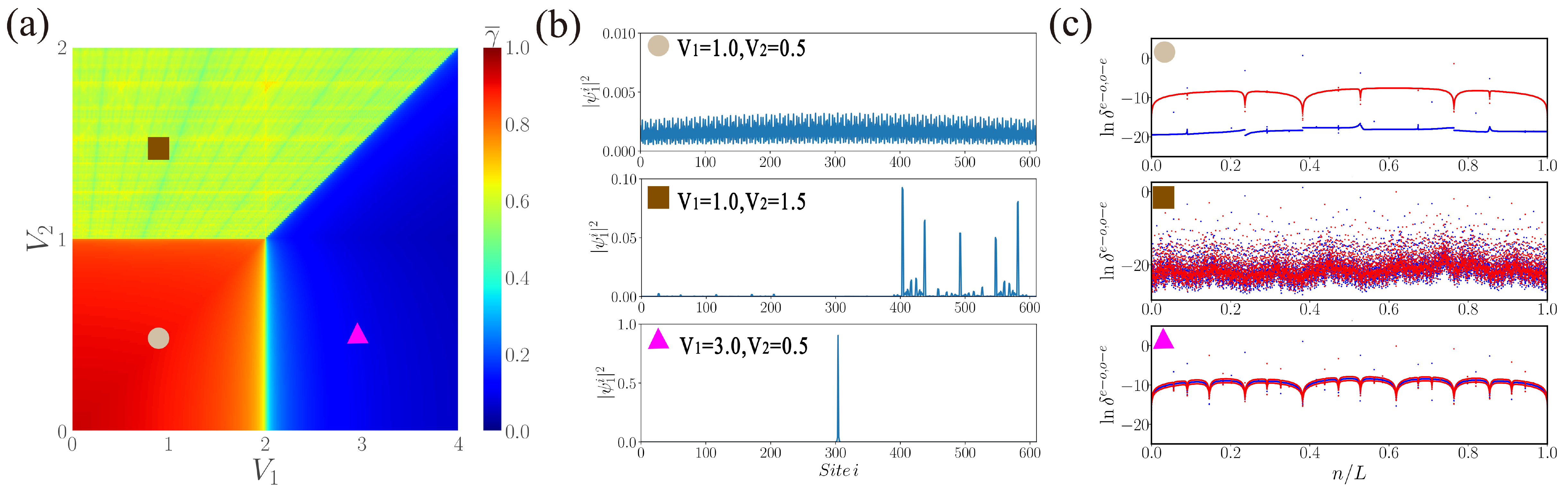}
\caption{The localized and critical properties of the system (\ref{eq1}) with $\Delta = 0 $. (a) Phase diagram of the system, where the red regime denotes the extended phase, green regime denotes the critical phase and blue regime denotes the localized phase. This phase diagram is obtained by calculating mean fractal dimension $\bar{\gamma}$ (see the text for detail). The skew phase boundary is determined by $V_1=2V_2$. (b) Density distribution $|\psi_n^j|^2$ of ground state($n=1$). The images from top to bottom correspond to extended, critical and localized state. Here, the system size is $L=610$. (c) The even-odd $\delta^{e-o}$ (red) and odd-even $\delta^{o-e}$ (blue) level spacings for the system size $L=17711$. The images from top to bottom correspond to extended, critical and localized phase in (a), where points of different colors are used to mark and correspond to the phase in (a).}
\label{fig1}
\end{figure*}
This paper is organized as follows. In section \ref{section2}, we introduce the generalized AAH model and its Hamiltonian.
In section \ref{section3}, we introduce the measurements, such as the inversed participation ratio, normalized participation ratio and fractal dimension.
In section \ref{section4}, the phase diagrams of the system and the multiple localization transitions are studied.
Based on the finite-size analysis of eigenstates in the intermediate phase, we obtain
a quantum phase where the localized, extended and critical states are coexistent in the thermodynamic limit, which is explained in section \ref{section6}.
In section \ref{section7}, we study the dynamic evolution of some initial states.
The summary of main results and concluding remarks are presented in section \ref{section8}.

\section{The system}\label{section2}

The generalized AAH model considered in this paper is described by the Hamiltonian
\begin{equation}
    H=\sum_{j=1}^L[t_j(c_{j+1}^{\dagger}c_j+h.c.)+ (\lambda_j +(-1)^j\Delta) n_j].   
	\label{eq1}
\end{equation}
Here $c^{\dagger}_j$ and $c_j$ are the fermionic creation and annihilation operators at $j$-th site, respectively.
$L$ is the system size, which is chosen as the Fibonacci number. Thus the critical states in this quasiperiodic system has certain fractal structures.
$n_j=c^{\dagger}_j c_j$ is the particle number operator.
$t_j=t+V_j$ quantifies the nearest-neighbor hopping, and $t = 1$ as energy unit. $V_j=V_2\cos[2\pi(j+1/2)\alpha+\theta]$ is the off-diagonal hopping, $V_2$ is the hopping amplitude,
$\alpha$ is an irrational number. In this paper, we chose $\alpha=\lim_{m\rightarrow\infty}\frac{F_{m-1}}{F_m}=\frac{\sqrt{5}-1}{2}$ and $F_m$ is the $m$-th Fibonacci number defined recursively by $F_m = F_{m-2}+F_{m-1}$ and $F_0 = F_1 = 1$.
$\lambda_j=V_1\cos(2\pi j \alpha+\theta)$, where $V_1$ and $\theta$ are the modulation amplitude and phase factor, respectively.
It is clear that the on-site potential is staggered due to the existence of $(-1)^j\Delta$ and $\Delta$ is the strength.
In general, the phase factor $\theta$ is the random numbers in the interval $[0,2\pi)$.
The boundary condition of the system (\ref{eq1}) is the periodic one.

The model (\ref{eq1}) has following generations.

(i) If $V_2=0$ and $\Delta=0$, the model (\ref{eq1}) is reduced to the AAH model. The system is in the extended phase if $V_1$ is small
and is in the localization phase if $V_1$ is big. The localization transition happens at the critical point of $V_1=2$.
The system does not have the single particle mobility edge thus the intermediate phase.

(ii) If $\Delta=0$, the model (\ref{eq1}) is reduced to the AAH model with off-diagonal hopping and on-site quasiperiodic potential.
The phase diagram of the system contains three phases: extended, localized and critical ones \cite{chang1997multifractal,takada2004statistics,liu2015localization,tang2021localization,wang2021many,zhai2020many}.

%\begin{figure*}
%\includegraphics[width=0.4\textwidth]{Fig1.pdf}

\section{The measurements}\label{section3}

\begin{figure*}[htb]
\includegraphics[width=1.0\textwidth]{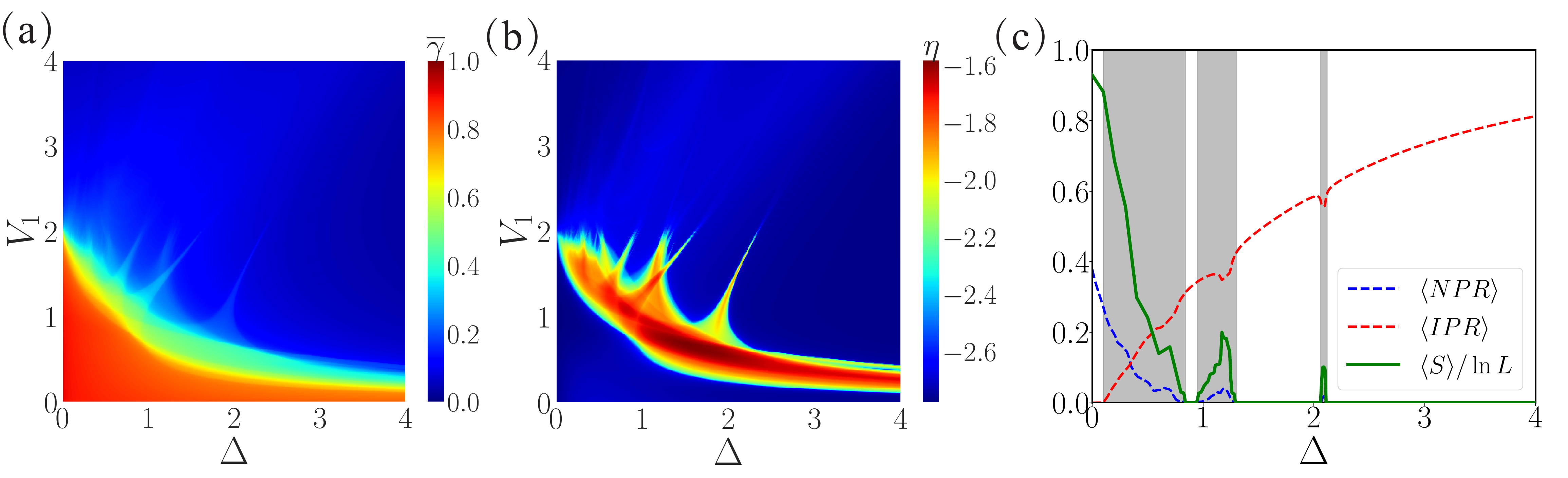}
\caption{(a) Phase diagram of the system (\ref{eq1}) in the $\Delta-V_1$ plane with fixed $V_2=0.5$, where the red regions denote the extended phase, green region denote the intermediate phase and blue region denote the localized phase. This phase diagram is also obtained by calculating mean fractal dimension $\bar{\gamma}$ (see the text for detail). (b) Complement to phase diagram of (a) by calculating $\eta$, which can distinguish the intermediate phase clearly. The blue regions represent the extended and localized phase, while other regions represent the intermediate phase. In (a) $\&$ (b), the system size is $L=610$. (c) The extrapolated values $\langle{\rm  IPR }\rangle$ (dashed red), $\langle {\rm NPR} \rangle$ (dashed blue) by calculating system size $L=1597, 2584, 4181, 6765$ and $\langle S \rangle / \ln L$ (solid green) for $L = 17711$ versus staggered onsite potential $\Delta$, where $V_1=1.5$, $V_2=0.5$. From the values of $\langle{\rm  IPR }\rangle$, $\langle {\rm NPR} \rangle$ and $\langle S \rangle / \ln L$, we see that the initial phase with $\Delta=0$ is extended while the final phase is localized, and the localization transitions happen three times with the increasing of $\Delta$. Here, the grey boxes mark intermediate phases.}
\label{fig2a}
\end{figure*}
In this section, we introduce several observable physical quantities to distinguish the extended, critical, localized states and the corresponding phases.
The first typical measurements are the inverse participation ratio ($\rm IPR $) and the corresponding fractal dimension ($\rm FD$). For a given single particle normalized eigenstate $\psi_n$, we can use the quantities \cite{li2020mobility}
\begin{equation}\label{ipr}
    {I}_n(q) = \sum_j |\psi_n^j|^{2q} \propto L^{-\gamma_{n}(q)},
\end{equation}
to characterize the details information of the eigenstate. Here, $\psi_n^j$ is the $j$-th element of $\psi_n$ and $\gamma_{n}(q)=D_n(q)(q-1)$. In our calculation, we choose $q = 2$ and the inverse participation ratio ${\rm IPR_{n}}={I}_n(2)$ and the fractal dimension       $ \gamma_{n}=D_n(2)$. $\rm IPR_{n}$ and $\gamma_{n}$ take difference values of the different regions in the large $L$ limit: 
${\rm IPR}_n$ tends to $1/L$ and $\gamma_{n}=1$ if $\psi_n$ is extended,
${\rm IPR}_n$ tends to $1$ and $\gamma_{n}=0$ if $\psi_n$ is localized and $0<\gamma_{n}<1$ and ${\rm IPR}_n$ tends to $L^{-\gamma_{n}}$ if $\psi_n$ is critical. The quasiperiodic system (\ref{eq1}) also has the critical states which are extended but non-ergodic. In order to characterize the critical states, we need to introduce the fractal dimension $\gamma_{n}$ of an eigenstate.
From the Eq.(\ref{ipr}) and set $q = 2$, we can get
\begin{equation}
-\ln({\rm IPR_{n}})/\ln(L)=-c/\ln(L)+\gamma_{n},
\end{equation}
where c is a size-independent coefficient. We can extrapolate the $\gamma_{n}$ by the intercept of the curve in the space spanned by $1/\ln(L)$ and $-\ln({\rm IPR_{n}})/\ln(L)$. For a large size system, we can simply ignore $-c/\ln(L)$ and get 
\begin{equation}
\gamma_{n} = -\ln({\rm IPR_{n}})/\ln(L).
\end{equation}
With the help of the fractal dimension $\gamma_{n}$, it is easy to determine the detailed states in the phases of the system.Taking the average of all the $\{\gamma_n\}$, we obtain the mean fractal dimension $\overline{\gamma}$
\begin{equation}
    \bar{\gamma}=\frac{1}{L}\sum_{n=1}^L \gamma_n,
\end{equation}
which can be used to distinguish the different phases. The system is in the extended phase if $\overline{\gamma}= 1$, in the
localized phase if $\overline{\gamma}= 0$, and in the intermediate or critical phase if $0<\overline{\gamma}<1$. When $0<\overline{\gamma}<1$, exactly what phase it belong to be, we need to further analysis energy spectrum with $\gamma_n$.
If all the eigenstates $\{\psi_n\}$ are critical, the system is in the critical phase.

These localization transitions are further complemented by inspecting the behavior of other parameters of interest such as the Shannon entropy and the normalized participation ratio ($\rm NPR$). The Shannon entropy is defined from a single particle state as $S_n = -\sum_j |\psi_n^j|^{2}\ln |\psi_n^j|^{2}$ \cite{padhan2021emergence,li2016quantum,sarkar2021mobility}, which vanishes for the localized states due to participation from a single site only and approaches its maximum value $\ln(L)$ for the extended states where the wave amplitude is finite for all lattice sites.  The  $\rm NPR$ is written as ${\rm NPR}_n=(L\sum_j |\psi_n^j|^{4})^{-1}$. Taking the average of all the $\{{\rm IPR}_n\}$ and that of $\{{\rm NPR}_n\}$, we obtain
\begin{equation}
\langle {\rm IPR} \rangle=\frac{1}{L}\sum_{n=1}^L{{\rm IPR}_n}, \quad \langle {\rm NPR }\rangle=\frac{1}{L}\sum_{n=1}^L{{\rm NPR}_n}.
\end{equation}
Then we conclude that in the thermodynamic limit where $L$ tends to infinity,
the system is in the extended phase if $\langle{\rm IPR}\rangle \simeq 0$ and $\langle{\rm NPR}\rangle$ is finite,
in the localized phase if $\langle{\rm IPR}\rangle$ is finite and $\langle{\rm NPR}\rangle \simeq 0$, and in the intermediate phase if both $\langle {\rm IPR} \rangle$ and $\langle {\rm NPR} \rangle$ are finite. We rely on the $\langle{\rm IPR}\rangle$ and  $\langle {\rm NPR} \rangle$ and obtain the phase diagrams by computing a introduced quantity $\eta$ \cite{li2020mobility,roy2021reentrant,padhan2021emergence}, which is defined as
\begin{equation}
\eta={\rm log}_{10}[\langle {\rm IPR} \rangle \times \langle {\rm NPR} \rangle].
\end{equation}
For our calculation, we set system size $L= 610$. When $\langle {\rm IPR} \rangle $ and $\langle {\rm NPR} \rangle$ $\sim \mathcal{O}(1) $, we get $-2.4 \lesssim \eta \lesssim -1.0$ in the intermediate phase. When one of them is close to $1/L$, we get $\eta \lesssim -\log_{10} L$ as $L \sim 10^3/2$, so $\eta \lesssim -2.5$ in extended and localized phases. We can use the quantity $\eta$ to clearly distinguish the intermediate region from the fully extended or the fully localized regions in the phase diagram.

In order to distinguish the extended, critical and localized states more clearly, we can define the even-odd (odd-even) level spacings of the eigenvalues as $\delta_{n}^{e-o}=E_{2n}-E_{2n-1}\left(\delta_{n}^{o-e}=E_{2n+1}-E_{2n}\right)$ \cite{deng2019one,padhan2021emergence,zhang2022localization,sarkar2021mobility}. $E_{2n}$ and $E_{2n-1}$ denote the even and odd eigenenergy in ascending order of the eigenenergy spectrum, respectively. In the extended states, the eigenengry spectrum for system is nearly doubly degenerate and cause $\delta_{n}^{e-o}$ to vanish. Hence there is an obvious gap between $\delta_{n}^{e-o}$ and $\delta_{n}^{o-e}$. In the localized states, $\delta_{n}^{e-o}$ and $\delta_{n}^{o-e}$ are almost same and the gap no longer exists. In the critical states, $\delta_{n}^{e-o}$ and $\delta_{n}^{o-e}$ have scatter-distributed behavior, which are different with extended and localized phases. Our results demonstrate that the different distribution of eigenvalues can be utilized to distinguish the different phases of the system (\ref{eq1}). All these quantities together confirm the multiple localization transitions and the existence of novel phase with extended, critical and localized states.

\section{Phase diagrams and multiple localization transitions}\label{section4}
Now, we are ready to calculate the phase diagram of the system (\ref{eq1}). Because the effect of $\theta$ in the large size system can be ignored, we consider the case of $\theta=0$ for the convenience of calculation.
Thus the system (\ref{eq1}) has three free model parameters $\Delta$, $V_1$ and $V_2$.
The phase diagram can be studied in the $\Delta-V_1$ plane with fixed $V_2$
and in the $\Delta-V_2$ plane with fixed $V_1$.

The phase diagram of the system (\ref{eq1}) with $\Delta=0$ is shown in Fig.\ref{fig1}(a).
From it, we see that the system has three phase: extended, localized and critical ones. There is only one type of state in each phase. For example, all eigenstates are critical in critical phase. In order to more intuitively see the difference between critical, extended and localized states, we plot the density distribution of the ground state corresponding to different phases at system size $L=610$ in Fig.\ref{fig1}(b). We also show the even-odd $\ln \delta^{e-o}$ (blue) and odd-even $\ln \delta^{o-e}$ (red) level spacings in Fig.\ref{fig1}(c) corresponding to different phases in Fig.\ref{fig1}(a) for the system size $L = 17711$. We can find the level spacing distribution of the critical phase is scattered in the middle of Fig.\ref{fig1}(c). For the extended phase there exists a gap between $\ln \delta^{e-o}$ and odd-even $\ln \delta^{o-e}$. For the localized phase, the gap vanishes.

There are four cases of localization transitions in Fig.\ref{fig1}(a).
(i) Along the line $V_2=0.5$, there has a transition from extended to localized phases, where the critical point is $V_1=2$.
(ii) Along the line $V_2=1.5$, there has a transition from critical to localized phases at the critical point of $V_1=3$.
(iii) Along the line $V_1=1$, there has a transition from extended to critical phases at the critical point of $V_2=1$.
(iv) Along the line $V_1=3$, there has a transition from localized to critical phases at the critical point of $V_2=1.5$.
According to these observations, the values of $V_2$ in the phase diagram of the system (\ref{eq1}) in the $\Delta-V_1$ plane are chosen as $0.5$ and $1.5$,
and the values of $V_1$ in the $\Delta-V_2$ plane are chosen as $1$ and $3$.

\begin{figure}[tbp]
\includegraphics[width=1.0\columnwidth]{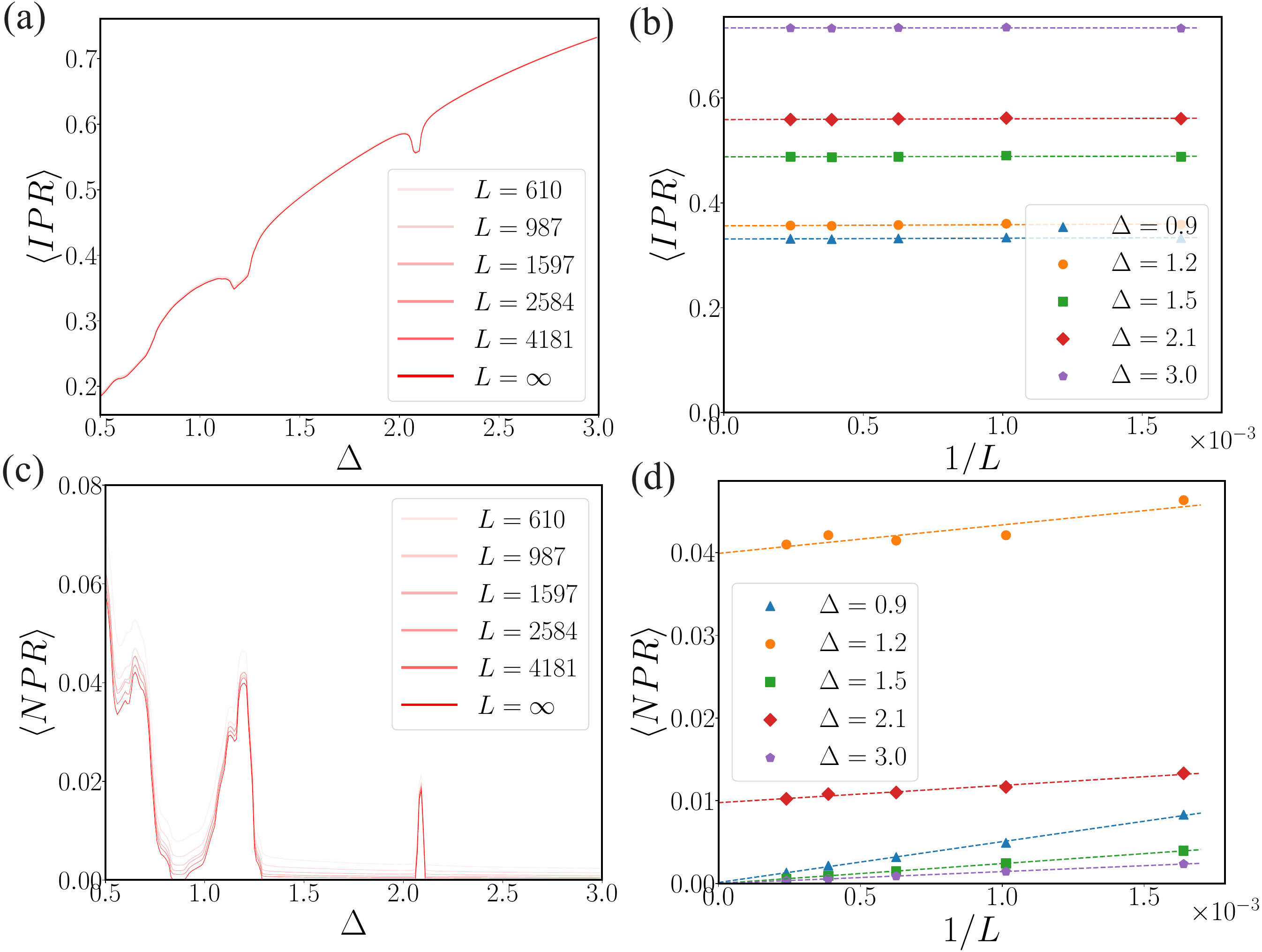}
\caption{(a) The $\rm \langle IPR \rangle$ for different system sizes such as $L=610,987,1597,2584,4181$ including $L=\infty$ (light to deep red curves) when $V_1=1.5$ and $V_2 = 0.5$. (b) Finite size extrapolation of $\rm \langle IPR \rangle$ as a function of $1/L$for some selected values of $\Delta$. (c) The $\rm \langle NPR \rangle$ for different system sizes including $L=\infty$ (light to deep red curves). (d)  Finite size extrapolation of $\rm \langle NPR \rangle$ as a function of $1/L$for some selected values of $\Delta$. }
\label{fsc}
\end{figure}
\begin{figure}[tbp]
\includegraphics[width=1.0\columnwidth]{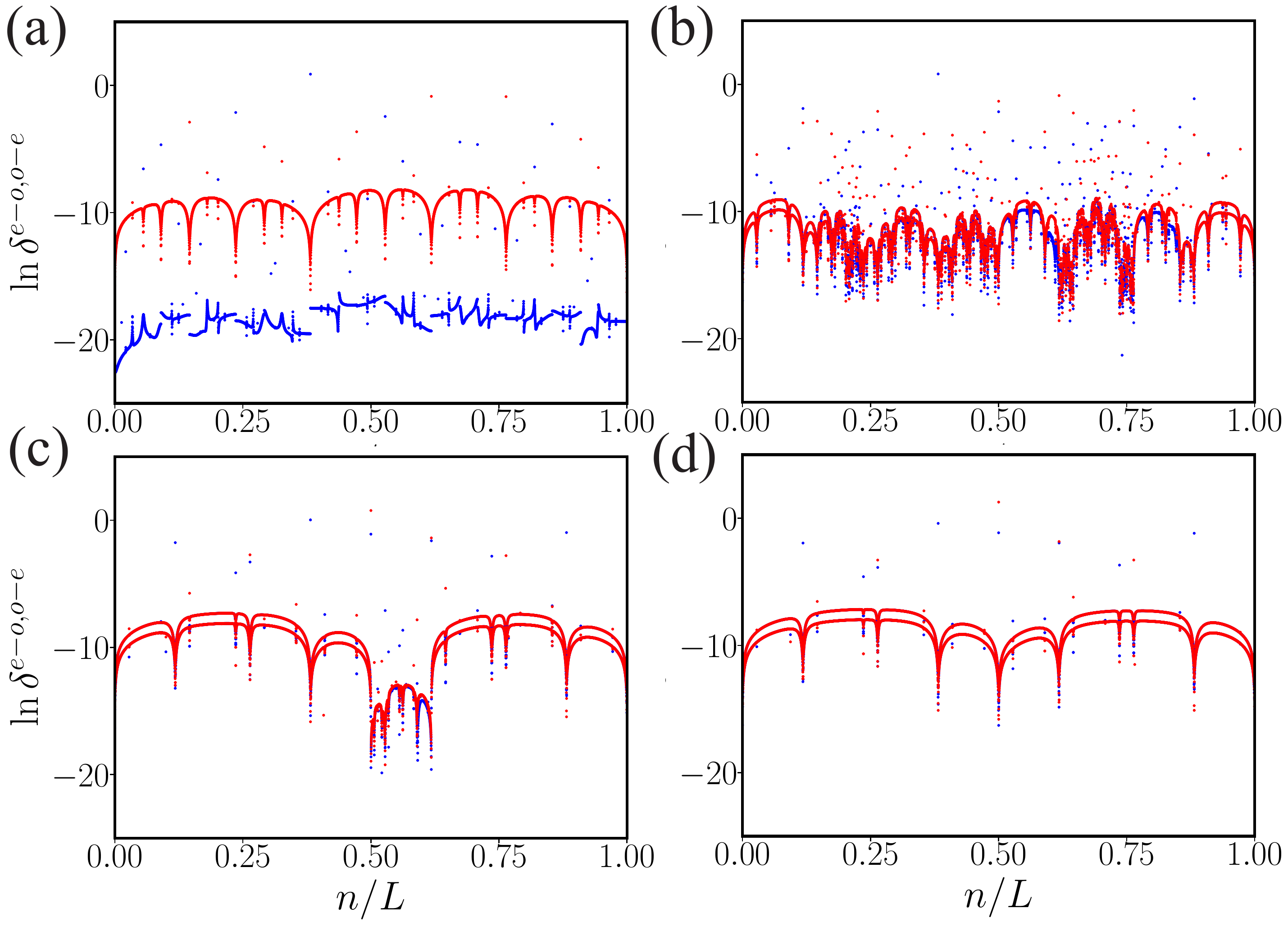}
\caption{The even-odd $\ln \delta^{e-o} $(blue) and odd-even $\ln \delta^{o-e}$(red) level spacings as a function of $n/L$ for the system with different staggered onsite potential $\Delta = 0$ (a), $0.4$ (b), $2.1$ (c) and $3$ (d). Here, $V_1 = 1.5,V_2=0.5$ and the system size $L = 17711$. (a) $\&$ (d) represent the case of fully expanded and localized phase, respectively. (b) $\&$ (c) represent the case of the intermediate phase, where the difference is that there exist critical states in (b), but not in (c). }
\label{eooe1}
\end{figure}
\subsection{Phase diagram in the $\Delta - V_1$ plane}
We first study the phase diagram of the system (\ref{eq1}) in the $\Delta-V_1$ plane with fixed $V_2=0.5$.
Based on the analyses of mean fractal dimension $\overline{\gamma}$ and $\eta$, we obtain the phase diagram, which is shown in Fig.\ref{fig2a}(a) and Fig.\ref{fig2a}(b). Compared with Fig.\ref{fig2a}(a), the existence of the intermediate phase can be seen more clearly in Fig.\ref{fig2a}(b).
We see that the system has three phases: extended, intermediate and localized ones. In Fig. \ref{fig2a}(b), the blue regions represent the extended and localized phase, while the other regions
represent the intermediate phase.

In the regime of $V_1>2$, the system is in the localized phase for any values of staggered on-site potential $\Delta$. Thus there is no the localization transition.

In the regime of $V_1<2$, the initial phase where $\Delta=0$ is extended. With the increasing of $\Delta$, the phase changes from extended to intermediate, then to localized. For some values of $V_1$ such as they are small, the localization transition happens once.
The most interesting thing is that for some intermediate values of $V_1$, the localization transition can be reentrant with the increasing of $\Delta$.
In order to show this phenomenon clearly, we plot the extrapolated value of $\rm \langle IPR \rangle$ and $\rm \langle NPR \rangle$ and $\langle S \rangle / \ln L$ versus $\Delta$ with the fixed $V_1=1.5$. The extrapolation of $\rm \langle IPR \rangle$ and $\rm \langle NPR \rangle$ will be introduced below. Here, $\langle S \rangle / \ln L$ indicates to sum the $S_n$ of all eigenstates and average $\ln L$, which can control the value range from 0 to 1. For the extended phase and localized phase, $\langle S \rangle / \ln L$ tend to 1 and 0 in the large size system, while for the intermediate phase, $\langle S \rangle / \ln L$ is finite.  The results are given in Fig.\ref{fig2a}(c). We see that with the help of three intermediate phases (grey regions), the localization transition occurs three times.

Then we fix $\Delta$ and tune $V_1$. We find that when the given $\Delta$ is very large, the localization transition happens once.
The significant thing is that when the staggered on-site potential $\Delta$ is suitable, the localization transition can be reentrant with the increasing of $V_1$.

Next, we consider the phases of the system (\ref{eq1}) with fixed $V_2=1.5$.
In this case, the extended phase is missing and the initial phase with $\Delta=0$ is the critical one.
After inducing the staggered on-site potential $\Delta$, only the transition from intermediate phase to localized phase occurs. We find that the
$\Delta$ can decrease the critical values of $V_1$.

\begin{figure*}[tb]
\includegraphics[width=1.0\textwidth]{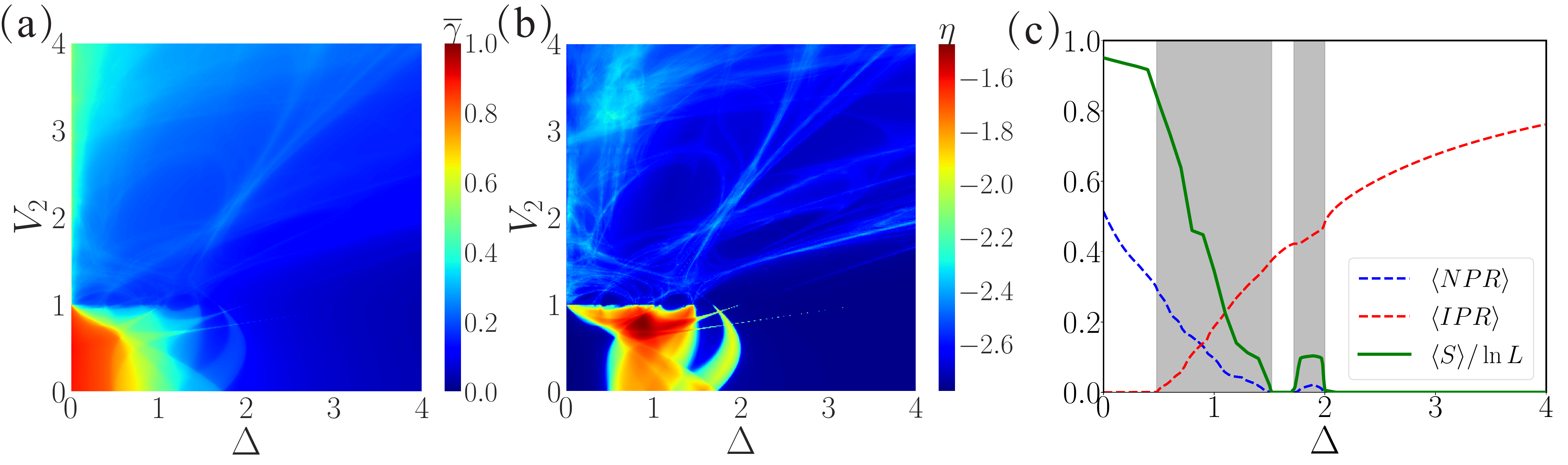}
\caption{(a) Phase diagram of the system (\ref{eq1}) in the $\Delta-V_1$ plane with fixed $V_1 = 1.0$, where the red regions denote the extended phase, green regions denote the intermediate phase and blue regions denote the localized phase. This phase diagram is also obtained by calculating mean fractal dimension $\bar{\gamma}$ (see the text for detail). (b) Complement to phase diagram of (a) by calculating $\eta$, which can distinguish the intermediate phase clearly. The blue regions represent the extended and localized phase, while other regions represent the intermediate phase. In (a) $\&$ (b), the system size is $L=610$. (c) The extrapolated values $\langle{\rm  IPR }\rangle$ (dashed red), $\langle {\rm NPR} \rangle$ (dashed blue) by calculating system size $L=610, 987, 1597, 2584, 4181$ and $\langle S \rangle / \ln L$ (solid green) for $L = 17711$ versus staggered onsite potential $\Delta$, where $V_1=1.0$, $V_2=0.5$. From the values of $\langle{\rm  IPR }\rangle$, $\langle {\rm NPR} \rangle$ and $\langle S \rangle / \ln L$, we see the localization transitions happen twice with the increasing of $\Delta$. Here, the grey boxes mark intermediate phases. }
\label{fig3a}
\end{figure*}

Here, we also have performed the finite-size analysis to confirm that the multiple localization transitions are not a finite-size effect. In Fig.\ref{fsc}(a) and (c), we compute $\rm \langle IPR \rm \rangle$ and $\rm \langle NPR \rangle$ for different system sizes $L$ such as $L = 610,987,1597,2584,4181$. We choose some special $\Delta$ values for different $L$ to fit and draw the curve of $\rm \langle IPR \rm \rangle$ and $\rm \langle NPR \rangle$ as a function of $1/L$ in Fig.\ref{fsc} (b) and (d), respectively. When $1/L$ tends to 0, we can deduce the $\rm \langle IPR \rangle$  and $\rm \langle NPR \rangle$ values for $L \rightarrow \infty$. In this way, we can derive $\rm \langle IPR \rangle$ and $\rm \langle NPR \rangle$ corresponding to $L \rightarrow \infty$ with different $\Delta$ by finite-size analysis. Then we plot the curve of $\rm \langle IPR \rangle$ and $\rm \langle NPR \rangle$ as a function of $\Delta$ for different system sizes including $L \rightarrow \infty$ in Fig.\ref{fsc} (a) and (c). We find this system indeed undergoes three localization transitions and have three intermediate phases with increasing $\Delta$ when $V_1 =1.5$ and $V_2 = 0.5$.

To further distinguish the different phases and understand clearly the behavior of the eigenstates in the intermediate phase, we also plot the even-odd $\ln \delta^{e-o}$ (blue) and odd-even $\ln \delta^{o-e}$ (red) level spacings in Fig.\ref{eooe1}. For the extended phase, there exists a gap between $\ln \delta^{e-o}$ and $\ln \delta^{o-e}$ shown as in Fig.\ref{eooe1} (a). For the extended phase, gap no longer exists in Fig.\ref{eooe1} (d). However, we find there exists extended, critical and localized states for the intermediate phase in the Fig. \ref{eooe1} (b) when $V_1 = 1.5$, $V_2 = 0.5$ and $\Delta = 0.4$. We find two level spacings
spectrum are scattered and induce there exists critical states when $n/L \in (0.2, 0.25)$. The extended states exist around $n/L \simeq 0.6$ and $n/L \simeq 0.8$. In Fig. \ref{eooe1} (c), we show there exist extended and localized states in the intermediate phase when $V_1 = 1.5$, $V_2 = 0.5$ and $\Delta = 2.08$. We can see there exists a gap around $n/L \simeq 0.6$, which means existence of extended states. 

\begin{figure}[htb]
\includegraphics[width=1.0\columnwidth]{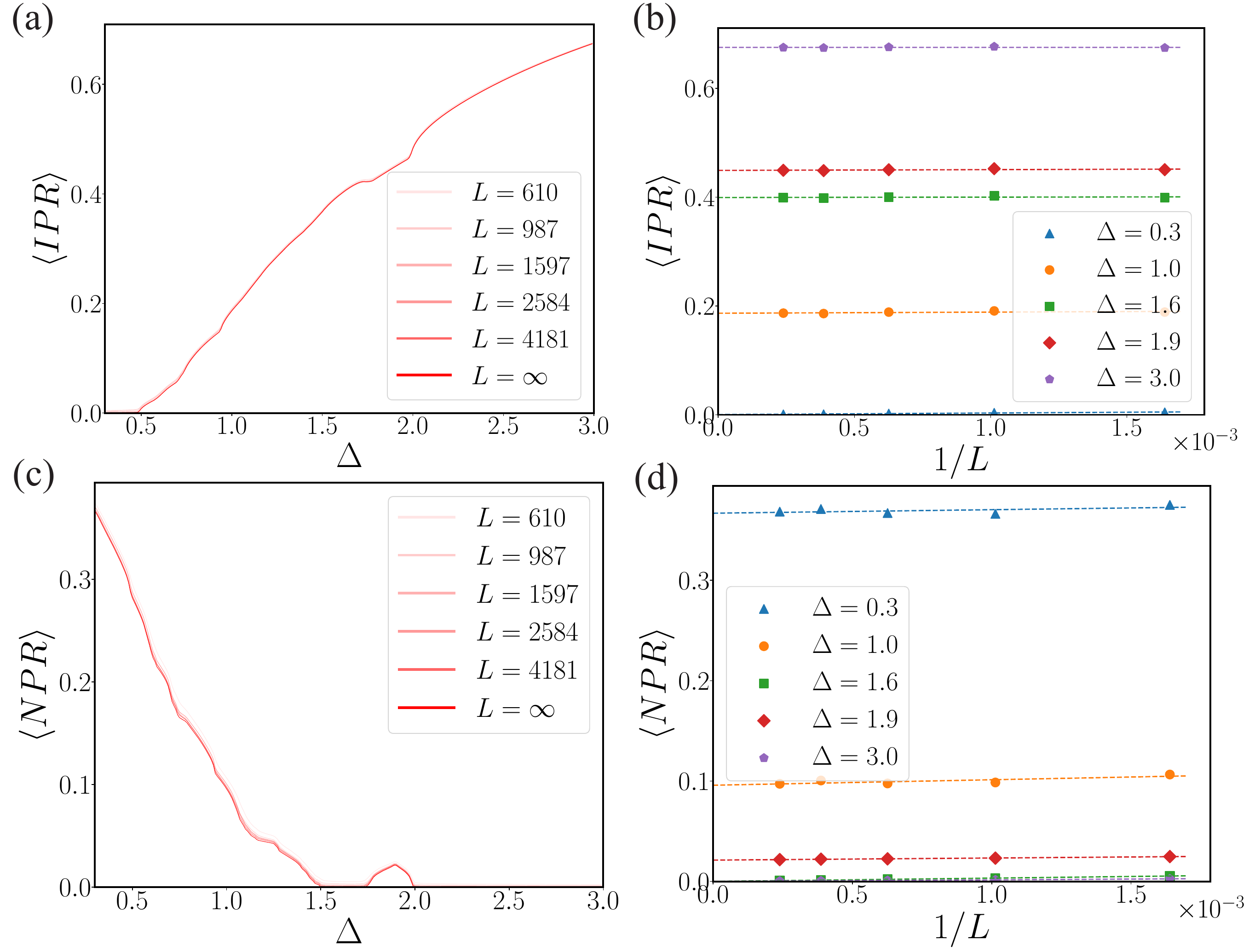}
\caption{(a) The $\rm \langle IPR \rangle$ for different system sizes such as $L=610,987,1597,2584,4181$ including $L=\infty$ (light to deep red curves) when $V_1=1.0$ and $V_2 = 0.5$. (b) Finite size extrapolation of $\rm \langle IPR \rangle$ as a function of $1/L$for some selected values of $\Delta$. (c) The $\rm \langle NPR \rangle$ for different system sizes including $L=\infty$ (light to deep red curves). (d)  Finite size extrapolation of $\rm \langle NPR \rangle$ as a function of $1/L$for some selected values of $\Delta$.}
\label{fig8}
\end{figure}

\subsection{Phase diagram in the $\Delta-V_2$ plane}
The phase diagram of the system (\ref{eq1}) in the $\Delta-V_2$ plane with fixed $V_1=1$ is shown in Fig.\ref{fig3a}(a) and (b).
We first analyze the regime of $V_2<1$. When the staggered on-site potential $\Delta$ is small, the system is in the extended phase.
With the increasing of $\Delta$, there exists the localization transition. In certain regimes of model parameters,
the localization transition can be reentrant.
For example, if $V_2=0.5$, from the values of $\rm \langle IPR\rangle$, $\rm \langle NPR\rangle$ and $\langle S \rangle/ \ln L$ given in Fig.\ref{fig3a}(c), we see that the localization transition happens twice.
We should note that if $\Delta$ is larger than $2$, the system is always in the localized phase.
Thus the localization transition and its reentrant occur only for the small $\Delta$.
\begin{figure}[b]
\includegraphics[width=1.0\columnwidth]{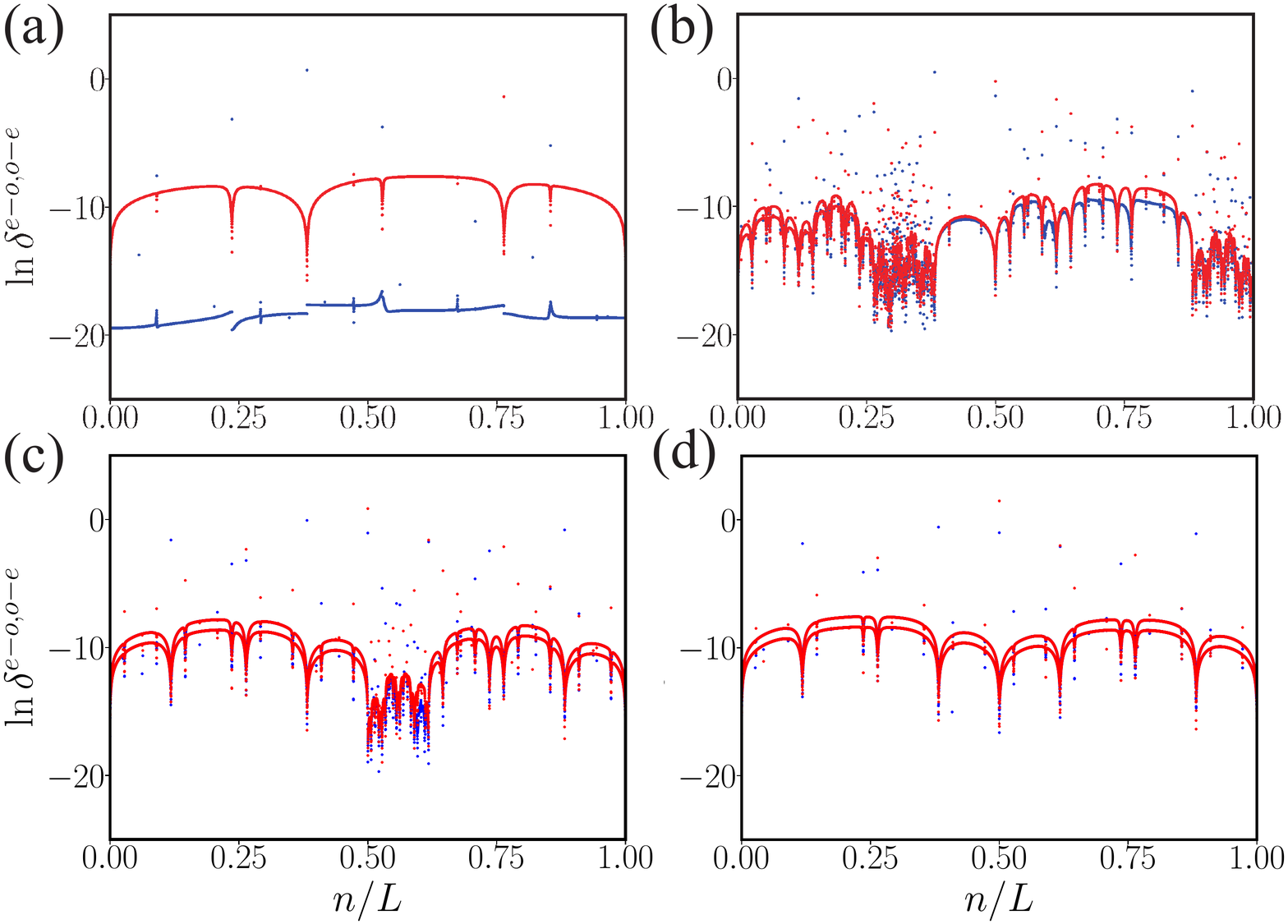}
\caption{The even-odd $\ln \delta^{e-o} $(blue) and odd-even $\ln \delta^{o-e}$(red) level spacings as a fuction of $n/L$ for the system with different staggered onsite potential $\Delta = 0$ (a), $0.7$ (b), $1.8$ (c) and $3.0$ (d). Here, $V_1 = 1.0,V_2=0.5$ and the system size $L = 17711$. (a) $\&$ (d) represent the case of fully expanded and localized phase, respectively. (b) $\&$ (c) represent the case of the intermediate phase, where the difference is that there exist critical states in (b), but not in (c).  }
\label{eooe2}
\end{figure}
\begin{figure*}[t]
\includegraphics[width=1.05\textwidth]{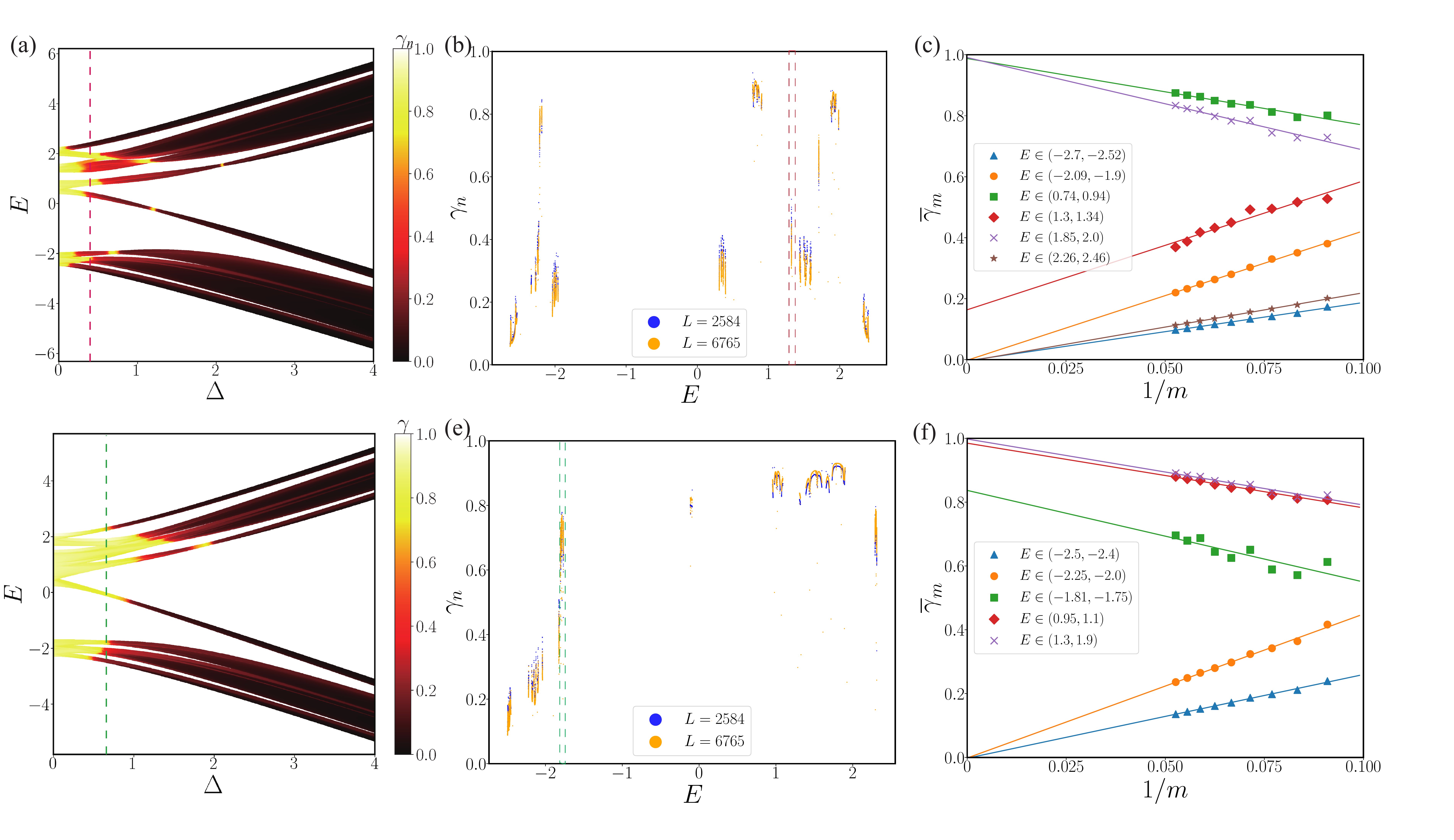}
\caption{ (a) $\&$ (d) The energy spectrum $E$ and fractal dimension $\gamma_n$ of each eigenstate of the system (\ref{eq1}) versus $\Delta$ with $L=610$, where (a) $V_1=1.5$, $V_2=0.5$ and (d) $V_1=1$, $V_2=0.5$. The red and green lines in (a) and (d) represent the two cases we consider respectively.
(b) $\&$ (e) The fractal dimension $\gamma_n$ of each eigenstate versus $E$ with (b) $V_1=1.5$, $V_2=0.5$, $\Delta=0.4$ and (e) $V_1=1.0$, $V_2=0.5$, $\Delta=0.7$ for different system size $L= 2584$ (blue) and $L = 6765$ (orange). The red and green boxes in (b) and (e) represent regions of critical states.
(c) $\&$ (f) The finite-size analysis of the mean fractal dimensions $\{\overline{\gamma}_m\}$ in different energy intervals, where (c) $V_1=1.5$, $V_2=0.5$, $\Delta=0.4$ and (f) $V_1=1.0$, $V_2=0.5$, $\Delta=0.7$. Here, we choose m to be equal 11 to 19. We see that the eigenstates in the energy intervals $(1.3,1.34)$ in figure (c) and $(-1.81,-1.75)$ in figure (f) are critical.}
\label{fig4}
\end{figure*}
In the regime of $V_2>1$, the initial intermediate phase with $\Delta=0$ is the critical one. 
With the increasing of $\Delta$, the intermediate phase transits to the localized phase. We find that with the increasing the staggered on-site potential, the critical value of $V_2$ at the transition point from extended phase to intermediate phase is decreased. Thus the critical states are sensitive to the staggered potential.
Here, we also perform finite-size analysis using same method to confirm that the reentrant transition is not a finite-size effect in Fig.\ref{fig8}. The relevant calculations are the same as in the previous section. We find this system indeed undergoes two localization transitions and have two intermediate phases with increasing $\Delta$ when $V_1 =1.0$ and $V_2 = 0.5$.

Similarly, we also plot the even-odd $\ln \delta^{e-o}$ (blue) and odd-even $\ln \delta^{o-e}$ (red) level spacings for system size $L = 17711$ in Fig.\ref{eooe2} same as Fig.\ref{eooe1}. For the extended phase, there exists a gap between $\ln \delta^{e-o}$ and $\ln \delta^{o-e}$ shown as in Fig.\ref{eooe2} (a) when $V_1=1.0$, $V_2 = 0.5$ and $\Delta =0 $. For the localized phase, gap no longer exists in Fig.\ref{eooe2} (d) when $V_1=1.0$, $V_2 = 0.5$ and $\Delta = 3 $. For intermediate phase, We also induce there exists extended, critical and localized regions in the Fig. \ref{eooe1} (b) when $V_1 = 1.5$, $V_2 = 0.5$ and $\Delta = 0.7$. We find there exists critical states when $n/L \in (0.25, 0.4)$. The extended states exist around $n/L \in (0.5, 0.8)$. In Fig. \ref{eooe2} (c), we show there exist extended and localized states in the intermediate phase when $V_1 = 1.0$, $V_2 = 0.5$ and $\Delta = 1.8$. We can see there exists a gap around $n/L \sim 0.6$, which means existence of extended states. 

\section{Coexistent phase with extended, critical and localized states}\label{section6}

The next task is that we should analyze the detailed states in the intermediate phases and check whether the critical phase survives when the staggered on-site potential is added.
For this purpose, we study the fractal dimension $\gamma_n$ of each eigenstate $\psi_n$.
As mentioned above, the fractal dimension $\gamma_n$ is finite for the critical state,
is zero for the localized state and is one for the extended state in the thermodynamic limit.
Here we use following method to calculate the limit behavior of $\gamma_n$ with $L\rightarrow \infty$ \cite{wang2022quantum,liu2022anomalous,roy2021critical}. The $\gamma_n$ is determined by the staggered parameter $\Delta$ thus the eigenenergy $E$.
We fist calculate the energy spectrum of the system. Based on them, we obtain the patterns of $\gamma_n$ versus $E$.
Please note that the system size $L$ is chosen as the $m$-th Fibonacci number $F_m$, and the patterns of $\gamma_n$ have certain fractal structures.
According to the patterns, we choose some small energy zones and calculate the mean fractal dimensions $\{\bar \gamma_m\}$ of these zones.
Obviously, the values of $\{\bar \gamma_m\}$ depend on the system size.
Thus we take the finite size scaling analysis of $\{\bar \gamma_m\}$ and obtain the values of $\{\bar \gamma_m\}$ in the thermodynamic limit.
We denoted the finial results as $\{\gamma_n\}$.
If $\gamma_n$ is finite, the corresponding eigenstates are critical.
If $\gamma_n=1$, the corresponding eigenstates are extend and if $\gamma_n=0$, the corresponding eigenstates are localized.

We first consider the intermediate phase shown in Fig.\ref{fig2a}(b), where the model parameters $V_1=1.5$, $V_2=0.5$ and $\Delta$ is free.
The energy spectrum and fractal dimension of each eigenstate versus $\Delta$ are shown in Fig.\ref{fig4}(a).
We see that the intermediate phase is not the critical phase, because the extended and localized states are included. Thus after inducing the
staggered potential $\Delta$, the critical phase is broken.

Usually, the intermediate phase of quasiperiodic system is a mixture of extended and localized states. Here we obtain that when the staggered potential $\Delta$ is suitable,
the intermediate phase can include the critical states, which is very rare.
Now we demonstrate this conclusion. We fix $\Delta=0.4$ and plot the curve of fractal dimension $\gamma_n$ of each eigenstate versus the eigenenergy $E$, which is shown in Fig. \ref{fig4}(b).
We see that the the fractal dimensions have some patterns. Meanwhile, the patterns move up or down with the increasing of system size $L$.
Choosing some small energy intervals, we calculate the mean fractal dimensions $\{\overline{\gamma}_m\}$.
The finite-size scaling behavior of $\{\overline{\gamma}_m\}$ is shown in Fig.\ref{fig4}(c), where
$1/m$ is the re-scaled system size. In the thermodynamic limit where $1/m\rightarrow 0$, we find that some of $\{\overline{\gamma}_m\}$ tend to 0, which correspond the localized states,
some of $\{\overline{\gamma}_m\}$ tend to 1, which correspond the extends states,
and that in the energy interval $(1.3,1.34)$ is finite, which means the eigenstates in this energy interval are critical.
Then we conclude that the system has a phase where the extended, localized and critical states are coexistent.
\begin{figure}[tbp]
\includegraphics[width=1.0\columnwidth]{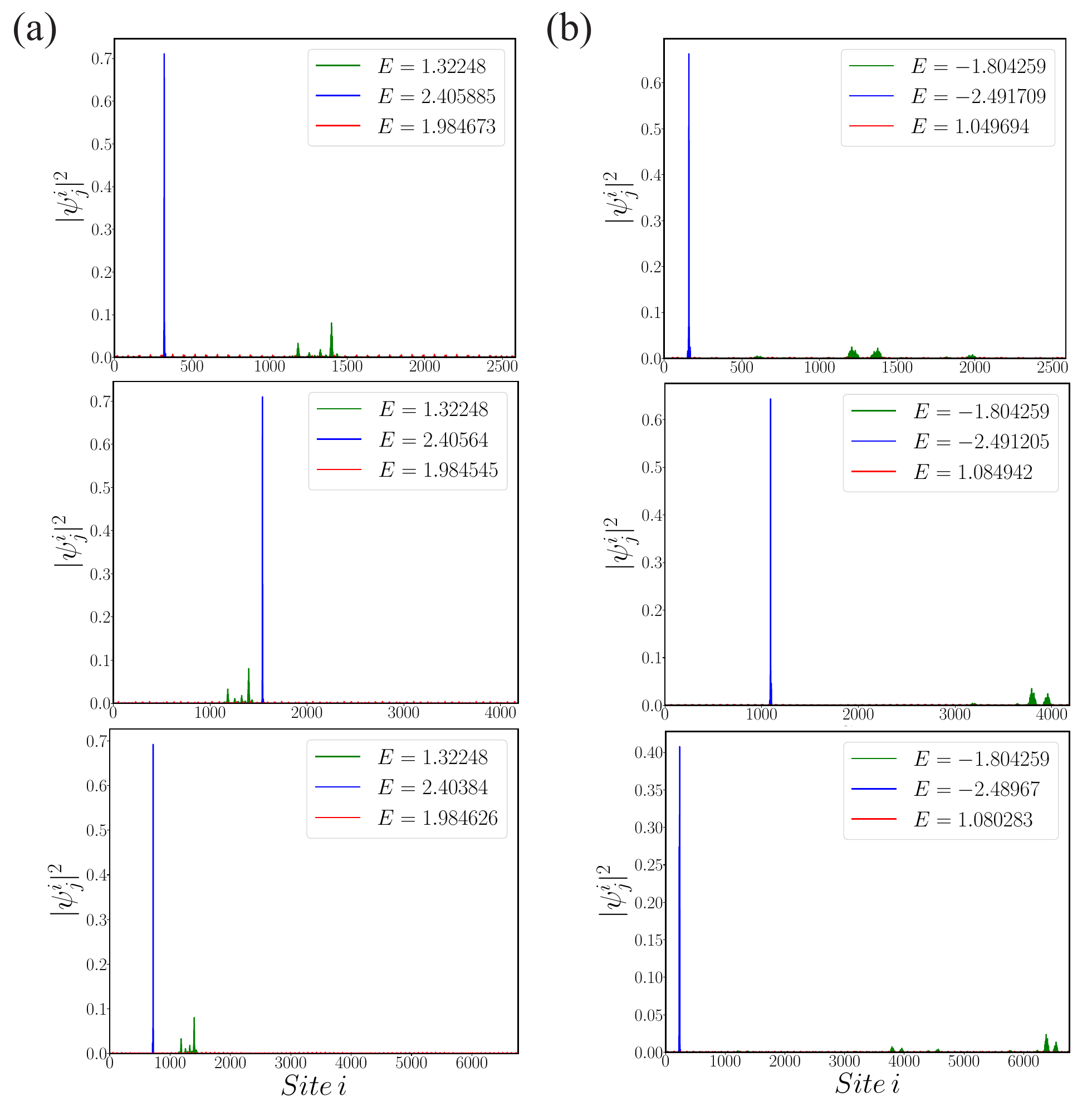}
\caption{Density distribution $|\psi_j^i|^2$ of different states in some energy regions for different system sizes $L=2584,4181,6765$. Here, we consider two case (a) $V_1=1.5,V_2 = 0.5, \Delta = 0.4$ and (b) $V_1=1.0,V_2=0.5,\Delta = 0.7$ and plot the density distribution of different states --- extended (red), critical (green) and localized (blue) states for different system sizes. The top is $L=2584$, the middle is $L=4181$ and the bottom is $L=6765$.}
\label{fig9}
\end{figure}

\begin{figure}[tbp]
\includegraphics[width=1.0\columnwidth]{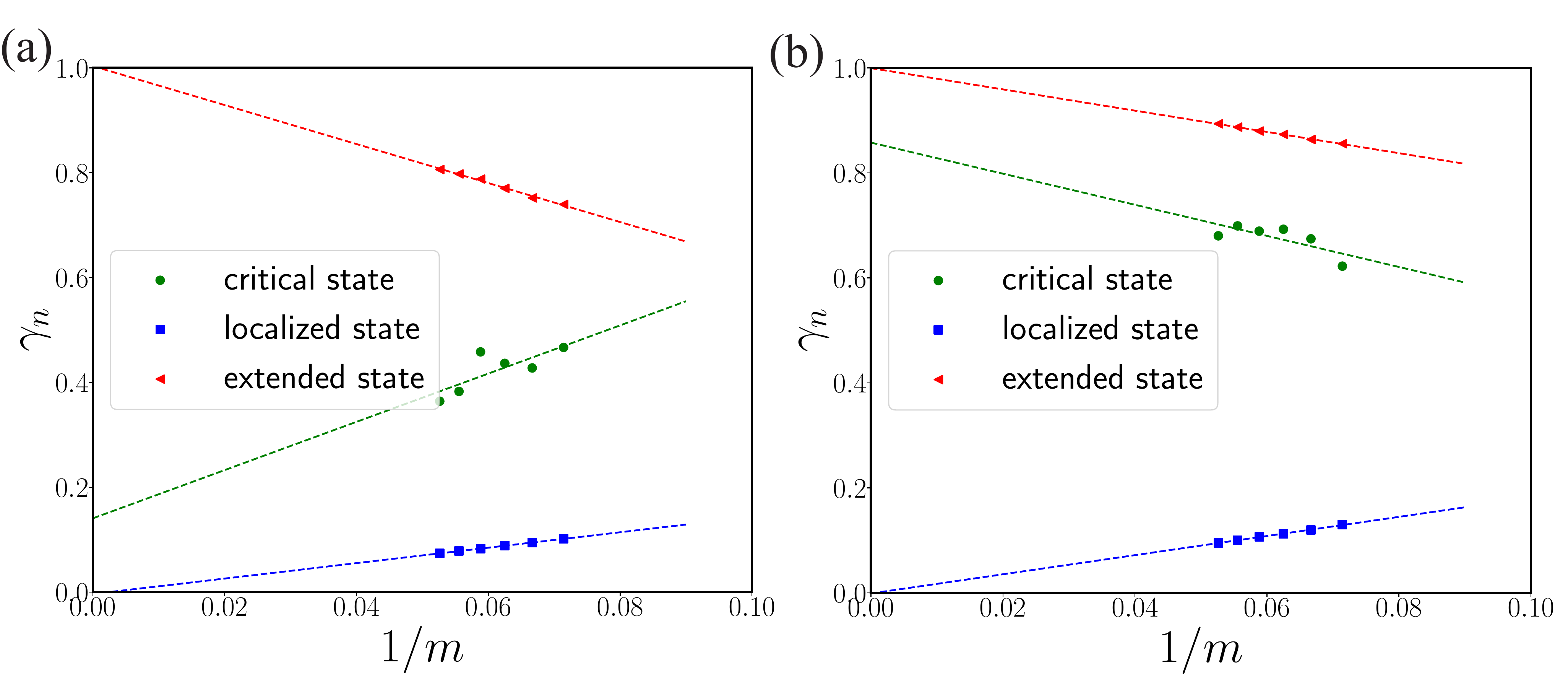}
\caption{The finite-size analysis of fractal dimensions $\gamma_n$ as a function of $1/m$ for the three states corresponding to ones in the top of the Fig.7(a) and (b), where system size $L=F_m$. Here, we consider two cases (a) $V_1=1.5,V_2 = 0.5, \Delta = 0.4$ and (b) $V_1=1.0,V_2=0.5,\Delta = 0.7$. The red line and dots represents the extended states. The green and blue represent the critical and localized states, respectively.}
\label{fig10}
\end{figure}
Next, we consider the intermediate phase shown in Fig. \ref{fig3a}(b).
The energy spectrum and fractal dimension of each eigenstate versus the staggered potential $\Delta$ are shown in Fig. \ref{fig4}(d).
The patterns of fractal dimensions with fixed $\Delta=0.7$ are shown in Fig. \ref{fig4}(e), and
the finite size scaling behavior of mean fractal dimensions $\{\overline{\gamma}_m\}$ in some energy intervals are shown in Fig. \ref{fig4}(f).
We see that the eigenstates in the energy interval $(-1.81,-1.75)$ are critical,
while the eigenstates in other intervals are either extended or localized.
Therefore, the critical states can be coexistent with the extended and localized states.
We shall note that this phenomenon is absent in the AAH model only with off-diagonal hopping.

In addition, in order to make our conclusion more convincing, we pick out three different states in different energy intervals, and draw the density distribution of the three different eigenstates at different system sizes such as $L = 2584, 4181,6765$ respectively in Fig. \ref{fig9}. For example, when $V_1=1.5$, $V_2=0.5$ and $\Delta =0.4$, we pick some cirtical states with the eigenenergy in the range $(1.3,1.34)$ in Fig. \ref{fig9}(a). Here, it is possible to choose a critical state with $E = 1.32248$ for different system sizes. We pick some extended and localized states with eigenenergy in the range $(1.85,2.0)$ and $(2.26,2.46)$. We find there is still a critical state with the system size increases. Similarly, when $V_1 = 1.0$, $V_2=0.5$ and $\Delta = 0.7$, we pick ciritical state with the eigenenergy $E = -1.804259$ in the range $(-1.81,-1.75)$ and draw the density distribution of different eigenstates in Fig. \ref{fig9}(b).

We also perform the finite-size analysis on the corresponding three states for the two cases we considered. We choose three eigenstates whose corresponding eigenenergy equal ones in the top of the Fig. \ref{fig9}(a) and (b). We calculate $\gamma_n$ of different states for different system size $L=F_m$, $m = 14,15,16,17,18,19$ in the Fig. \ref{fig10}. When $1/m\rightarrow 0$, it is possible to extrapolate $\gamma_n$  at the thermodynamic limit $L\rightarrow \infty$. We prove there exists three states --- extended ($\gamma_n = 1$), critical($0<\gamma_n<1$) and localized($\gamma_n$ = 0) states in this system at the thermodynamic limit as shown in Fig. \ref{fig10}(a) and (b).

\begin{figure}[tbp]
\includegraphics[width=1.0\columnwidth]{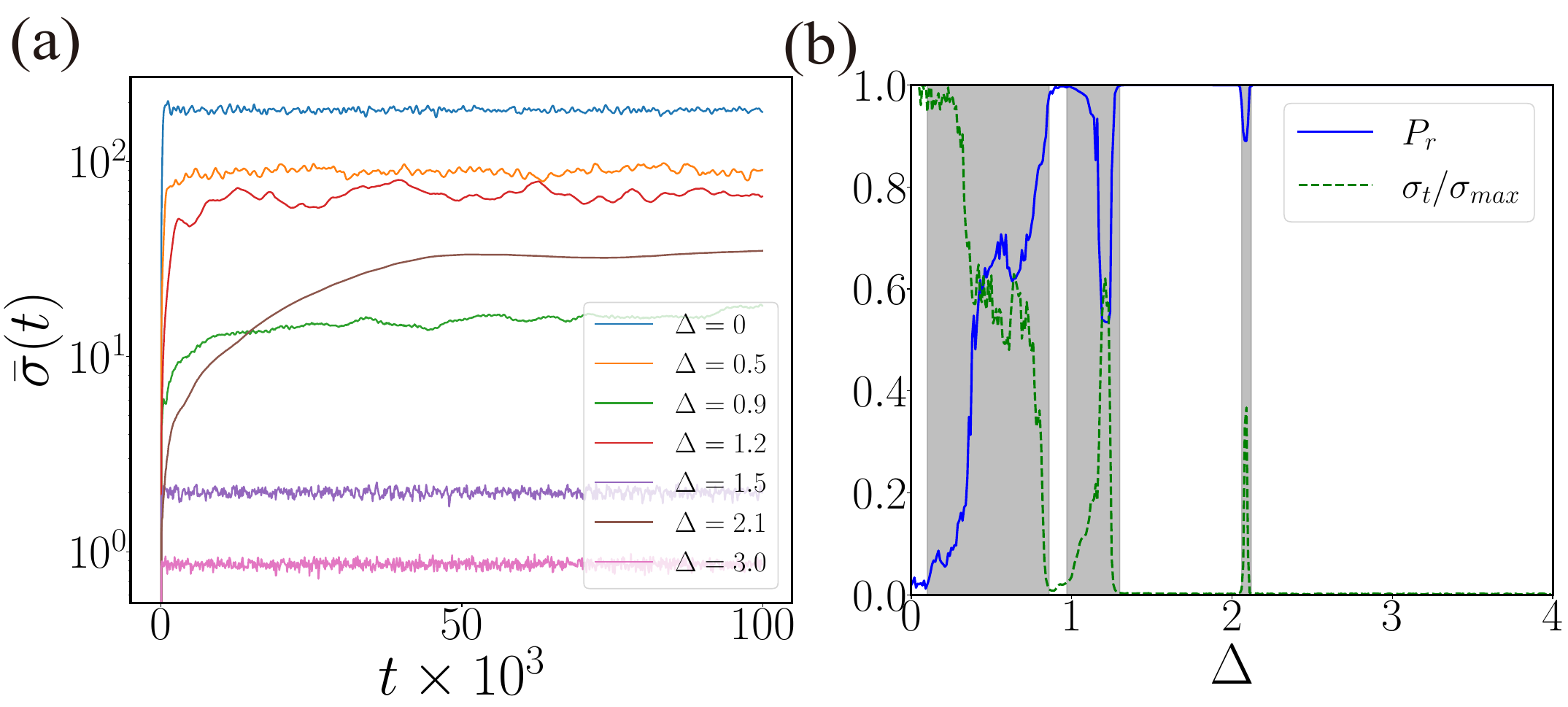}
\caption{(a) The time evolution of $\bar{\sigma}(t)$ with different staggered quasiperiodic potentials, where $L=610$, $V_1=1.5$ and $V_2=0.5$.
(b) The survival probability $P_r$ and $\sigma_t/\sigma_{max}$ at $t=10^5$ versus $\Delta$, where $L=2584$, $r=40$, $V_1=1.5$ and $V_2=0.5$. We see that there indeed exist the multiple localization transition, as given by Fig. \ref{fig2a}(b). Here, the grey boxes mark intermediate phases.}
\label{fig2}
\end{figure}

\section{Dynamic evolution}\label{section7}

In this section, we study the dynamic properties of the system (\ref{eq1}) with open boundary condition. The time evolution of a given initial state $|\Psi(0)\rangle$
is determined by
\begin{equation}
	|\Psi(t)\rangle=e^{-iHt}|\Psi(0)\rangle,
\end{equation}
where $H$ is given by Eq. (\ref{eq1}) and we have set $\hbar=1$.
Here the initial state is chosen as
$j_0$-th basis of the Hilbert space, $|\Psi(0)\rangle=|j_0\rangle$, i.e., a particle locates at the $j_0$-th site of the chain at the initial time. %$j_0=L/2$ if $L$ is even and $j_0=(L+1)/2$ if $L$ is odd.
Because the system (\ref{eq1}) is a single particle model, the state $|\Psi(t)\rangle$ can be decomposed as $|\Psi(t)\rangle=\sum_{j=1}^{L} \psi_j(t)|j\rangle$,
where $\psi_j(t)$ is the time-dependent wave function. With the help of $\psi_j(t)$,
a dynamic quantity named root mean-square displacement $\sigma(t)$ is proposed \cite{zhang2012quantum,xu2020dynamical}
\begin{equation}\label{eq9}
	\sigma(t)= \sqrt{\sum_{j=1}^{L}(j-j_0)^2|\psi_j(t)|^2}.
\end{equation}

Because the localized states don't diffuse in the long time evolution,
the saturation value of $\sigma(t)$ in the localized phase is smaller than those in the extended or intermediate phase.
Here, we also perform averaging over different initial states, i.e., a particle initialized on randomly chosen sites far enough from the chain boundaries, so that we get the mean value of the $\sigma(t)$, $\bar{\sigma}(t) = \langle \sigma(t) \rangle_{j_0}$, where $\langle...\rangle_{j_0}$ represents averaging over different initial states that a particle is randomly located at $j_0$-th site. Here, we choose $100$ different $j_0$ from the region $[L/3, 2L/3]$ to perform numerical calculations. The results show that the dynamical properties are independent of the choice of the initial localized states, but depend on the parameters of the system.%For simplicity of the numerical calculation, we specifically choose a wave packet initially located at the center of the chain to calculate the $\sigma_t/\sigma_{max}$ and $P_r(t)$ in Figs. \ref{fig2}(b) and \ref{fig3}(b).}
According to the Eq. (\ref{eq9}), we take $j_0 = L/2$ and denote the value of $\sigma(t)$ after a long time evolution as $\sigma_t$.
We consider the quantity $\sigma_t/\sigma_{max}$, where $\sigma_{max}$ is the value of $\sigma_t$ with ceratin model parameter in the extended phase.
Here, $\sigma_{max}=\sigma_t|_{\Delta=0}$.
Then $\sigma_t/\sigma_{max}$ can be used to distinguish the different phases.
$\sigma_t/\sigma_{max}$ tends to 1 in the extended phase, tends to 0 in the localized phase and is finite in the intermediate phase.

By using the wave function $\psi_j(t)$, another observable physical quantity named survival probability $P_r(t)$ is proposed \cite{zhang2012quantum,wu2021non}
\begin{equation}
	P_{r}(t)=\sum_{j=\lceil \frac{L}{2} \rceil -r}^{ \lceil \frac{L}{2} \rceil +r}|\psi_j(t)|^2,
\end{equation}
where $\lceil L/2 \rceil$ means the smallest integer not less than $L/2$, and $r$ is a small integer. Obviously,
after a long time evolution, if the system is in the extended phase, the survival probability $P_{r}$ tends to 0.
If the system is in the localized phase, $P_{r}$ tends to 1.
The $P_{r}$ is finite in the intermediate phase.

\begin{figure}[tbp]
\includegraphics[width=1.0\columnwidth]{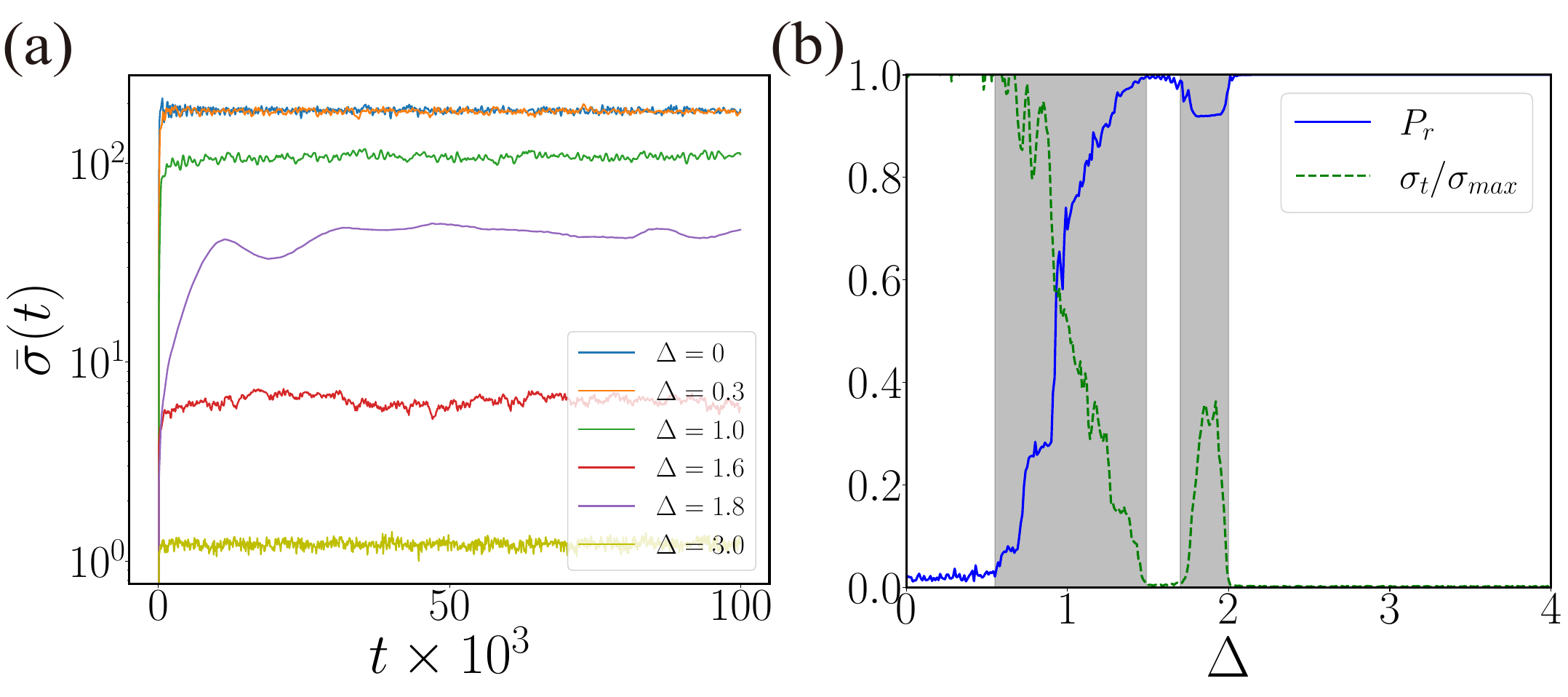}
\caption{(a) The time evolution of $\bar{\sigma}(t)$ with different $\Delta$, where $L=610$, $V_1=1.0$ and $V_2=0.5$.
(b) The survival probability $P_r$ and $\sigma_t/\sigma_{max}$ at $t=10^5$ versus $\Delta$, where $L=2584$, $r=40$, $V_1=1.0$ and $V_2=0.5$. We see that there indeed exist the multiple localization transition, as given by Fig. \ref{fig3a}(b). Here, the grey boxes mark intermediate phases.}
\label{fig3}
\end{figure}

The time evolutions of $\bar{\sigma}(t)$ with some fixed $\Delta$ are shown in Figs. \ref{fig2}(a) and \ref{fig3}(a).
The $\sigma_t/\sigma_{max}$ and $P_{r}(t)$ with $r=40$ at the time $t=10^5$ versus the $\Delta$ are shown in Figs. \ref{fig2}(b) and \ref{fig3}(b).
We see that there indeed exist the multiple localization transitions with the increasing of $\Delta$.These results are consistent with the ones obtained by $\rm \langle IPR \rangle$ and $\rm \langle NPR \rangle$.

\section{Summary}\label{section8}

In this paper, we have studied localization transitions and dynamical properties in the generalized AAH model with staggered on-site potential. Based on the analyses of $\rm \langle IPR \rangle$, $\rm \langle NPR \rangle$ and mean fractal dimension, we obtain the phase diagram of the system. We find that the critical phase is broken after inducing the staggered on-site potential. The system has the mobility edge, thus the extended and localized phases are separated by the intermediate phase. Interestingly, the staggered on-site potential can induce the multiple localization transition phenomena. Most importantly, by using the energy spectrum, patterns of fractal dimensions and finite-size analysis, we obtain a novel quantum phase where the extended, localized and critical states are coexistent in some regimes of model parameters. We also study the dynamic evolution in different phases with the help of root mean-square displacement and survival probability. Our theoretical results may be experimentally simulated in the future \cite{li2022observation,wang2022quantum,shimasaki2022anomalous}. It is worth exploring whether the interacting quasiperiodic system may have reentrant many-body localization transitions or may find a novel many-body intermediate phase with coexisting extended, critical and localized states \cite{wang2021many,li2015many}.

\section*{Acknowledgments}
We thank Tapan Mishra and Ashirbad Padhan for fruitful discussions. The financial supports from National Key R$\&$D Program of China (Grant No. 2021YFA1402104),
National Natural Science Foundation of China (Grant Nos. 12074410, 12047502, 11934015 and 11947301) and Strategic Priority Research Program of the Chinese Academy of Sciences (Grant No. XDB33000000) are gratefully acknowledged.

%\bibliographystyle{iopart-num.bst}
%\bibliography{NHME_dimerizedAAmodel}

\begin{thebibliography}{99}	
\bibitem{anderson1958absence} P. W. Anderson, Absence of diffusion in certain random lattices, Phys. Rev. \textbf{109}, 1492 (1958).

\bibitem{abrahams1979scaling} E. Abrahams, P. W. Anderson, D. C. Licciardello and T. V. Ramakrishnan, Scaling theory of localization: Absence of quantum diffusion in two dimensions, Phys. Rev. Lett. \textbf{42}, 673 (1979).

\bibitem{lee1985disordered} P. A. Lee and T. V. Ramakrishnan, Disordered electronic systems, Rev. Mod. Phys. \textbf{57}, 287 (1985).          %disordered electronic systems

\bibitem{evers2008anderson} F. Evers and A. D. Mirlin, Anderson transitions, Rev. Mod. Phys. \textbf{80}, 1355 (2008).             %Anderson transition


\bibitem{harper1955single} P. G. Harper, Single Band Motion of Conduction Electrons in a Uniform Magnetic Field, Proc. Phys. Soc. A \textbf{68}, 874 (1955).   %Single Band Motion of Conduction Electrons in a Uniform Magnetic Field

\bibitem{aubry1980analyticity} S. Aubry and G. Andr{\'e}, Analyticity breaking and Anderson localization in incommensurate lattices, Ann. Israel Phys. Soc. \textbf{3}, 18 (1980).       %Analyticity breaking and Anderson localization in incommensurate lattices.
\bibitem{luschen2018single} H. P. L{\"u}schen, S. Scherg, T. Kohlert, M. Schreiber, P. Bordia, X. Li, S. D. Sarma and I. Bloch, Single-Particle Mobility Edge in a One-Dimensional Quasiperiodic Optical Lattice, Phys. Rev. Lett. \textbf{120}, 160404 (2018).

\bibitem{li2017mobility} X. Li, X. Li and S. D. Sarma, Mobility edges in one-dimensional bichromatic incommensurate potentials, Phys. Rev. B \textbf{96}, 085119 (2017).      %Mobility edges in one-dimensional bichromatic incommensurate potentials


\bibitem{sarma1988mobility} S. D. Sarma, S. He and X. C. Xie, Mobility Edge in a Model One-Dimensional Potential, Phys. Rev. Lett. \textbf{61}, 2144 (1988).     %Mobility Edge in a Model One-Dimensional Potential

\bibitem{sarma1990localization} S. D. Sarma, S. He and X. C. Xie, Localization, mobility edges, and metal-insulator transition in a class of one-dimensional slowly varying deterministic potentials, Phys. Rev. B \textbf{41}, 5544 (1990).      %Localization, mobility edges, and metal-insulator transition in a class of one-dimensional slowly varying deterministic potentials

\bibitem{biddle2010predicted} J. Biddle  and S. D. Sarma, Predicted Mobility Edges in One-Dimensional Incommensurate Optical Lattices: An Exactly Solvable Model of Anderson Localization, Phys. Rev. Lett. \textbf{104}, 070601 (2010).%Predicted Mobility Edges in One-Dimensional Incommensurate Optical Lattices: An Exactly Solvable Model of Anderson Localization

\bibitem{ganeshan2015nearest} S. Ganeshan, J. Pixley and S. D. Sarma, Nearest Neighbor Tight Binding Models with an Exact Mobility Edge in One Dimension, Phys. Rev. Lett. \textbf{114}, 146601 (2015).  %Nearest Neighbor Tight Binding Models with an Exact Mobility Edge in One Dimension



%Mobility edges in one-dimensional bichromatic incommensurate potentials
\bibitem{li2020mobility} X. Li  and S. D. Sarma, Mobility edge and intermediate phase in one-dimensional incommensurate lattice potentials, Phys. Rev. B \textbf{101}, 064203 (2020).      %   Mobility edge and intermediate phase in one-dimensional incommensurate lattice potentials

\bibitem{wang2020one} Y. Wang, X. Xia, L. Zhang, H. Yao, S. Chen, J. You, Q. Zhou and X. Liu, One-Dimensional Quasiperiodic Mosaic Lattice with Exact Mobility Edges, Phys. Rev. Lett. \textbf{125}, 196604 (2020). %One-Dimensional Quasiperiodic Mosaic Lattice with Exact Mobility Edges

\bibitem{gonccalves2022hidden} M. Gon\c{c}alves, B. Amorim, E. V. Castro and P. Ribeiro, Hidden dualities in 1D quasiperiodic lattice models, SciPost Phys. \textbf{13}, 046 (2022).
\bibitem{RGM2022} M. Gon\c{c}alves, B. Amorim, E. V. Castro and P. Ribeiro, Renormalization-Group Theory of 1D quasiperiodic lattice models with commensurate approximants, arXiv:2206.13549v3 (2022).

\bibitem{criticalMG2023} M. Gon\c{c}alves, B. Amorim, E. V. Castro and P. Ribeiro, Critical phase dualities in 1D exactly-solvable quasiperiodic models, arXiv:2208.07886v2 (2023).




\bibitem{goblot2020emergence} V. Goblot, A. {\v{S}}trkalj, N. Pernet, J. L. Lado, C. Dorow, A. Lema{\^\i}tre, G. L. Le, A. Harouri, I. Sagnes, S. Ravets, A. Amo, J. Bloch and O. Zilberberg, Emergence of criticality through a cascade of delocalization transitions in quasiperiodic chains, Nat. Phys. \textbf{16}, 832 (2020).

\bibitem{roy2021reentrant} S. Roy, T. Mishra, B. Tanatar and S. Basu, Reentrant localization transition in a quasiperiodic chain, Phys. Rev. Lett. \textbf{126}, 106803 (2021).


\bibitem{wu2021non} C. Wu, J. Fan, G. Chen and S. Jia, Non-Hermiticity-induced reentrant localization in a quasiperiodic lattice, New J. Phys. \textbf{23}, 123048 (2021).


\bibitem{jiang2021mobility} X. -P. Jiang, Y. Qiao and J. Cao, Mobility edges and reentrant localization in one-dimensional dimerized non-Hermitian quasiperiodic lattice, Chin. Phys. B \textbf{30}, 097202 (2021).

\bibitem{zhai2021cascade} L. J. Zhai, G. Y. Huang and S. Yin, Cascade of the delocalization transition in a non-Hermitian interpolating Aubry-Andr{\'e}-Fibonacci chain, Phys. Rev. B \textbf{104}, 014202 (2021).

\bibitem{padhan2021emergence} A. Padhan, M. Giri, S. Mondal and T. Mishra, Emergence of multiple localization transitions in a one-dimensional quasiperiodic lattice, Phys. Rev. B \textbf{105}, L220201 (2022).

\bibitem{han2022dimerization} W. Q. Han and L. W. Zhou, Dimerization-induced mobility edges and multiple reentrant localization transitions in non-Hermitian quasicrystals, Phys. Rev. B \textbf{105}, 054204 (2022).

\bibitem{wang2023fate} H. Wang, X. Zheng, J. Chen, L. Xiao, S. Jia and L. Zhang, fate of the reentrant localization phenomenon in the one-dimensional dimerized quasiperiodic chain with long-range hopping, Phys. Rev. B \textbf{107}, 075128 (2023).


\bibitem{vaidya2022reentrant} S. Vaidya, C. J\"org, K. Linn, M. Goh and M. C. Rechtsman, Reentrant delocalization transition in one-dimensional photonic quasicrystals, arXiv:2211.06047v1 (2023).
\bibitem{nair2023emergent} P. S Nair, D. Joy and S. Sanyal, Emergent scale and anomalous dynamics in certain quasi-periodic systems, arXiv:2302.14053v1 (2023).


\bibitem{liu2015localization} F. Liu, S. Ghosh, and Y.D. Chong, Localization and adiabatic pumping in a generalized Aubry-Andr{\'e}-Harper model, Phys. Rev. B \textbf{91}, 014108 (2015).


\bibitem{takada2004statistics} T. Yoshihiro, I. Kazusumi and Y. Masanori, Statistics of spectra for critical quantum chaos in one-dimensional quasiperiodic systems, Phys. Rev. E \textbf{70}, 066203 (2004).



\bibitem{chang1997multifractal} I. Chang, K. Ikezawa and M. Kohmoto, Multifractal properties of the wave functions of the square-lattice tight-binding model with next-nearest-neighbor hopping in a magnetic field, Phys. Rev. B \textbf{55}, 12971 (1997).

\bibitem{zhai2020many} L. J. Zhai, S. Yin and G. Y. Huang, Many-body localization in a non-Hermitian quasiperiodic system, Phys. Rev. B \textbf{102}, 064206 (2020).

\bibitem{wang2021many} Y. C. Wang, C. Cheng, X. J. Liu and D. P. Yu, Many-body critical phase: extended and nonthermal, Phys. Rev. Lett. \textbf{126}, 080602 (2021).

\bibitem{tang2021localization} L. Z. Tang, G. Q. Zhang, L. F. Zhang and D. W. Zhang, Localization and topological transitions in non-Hermitian quasiperiodic lattices, Phy. Rev. A \textbf{103}, 033325 (2021).

\bibitem{zhang2022localization} Y. Zhang, B. Zhou, H. Hu, and S. Chen, Localization, multifractality, and many-body localization in periodically kicked quasiperiodic lattices, Phys. Rev. B \textbf{106}, 054312 (2022).
\bibitem{wang2022mobility} Y. Wang, Mobility edges and critical regions in a periodically kicked incommensurate optical Raman lattice, Phys. Rev. A, \textbf{106}, 053312 (2022).

\bibitem{li2022observation} H. Li, Y. Y. Wang, Y. H. Shi, K. X Huang, X. H. Song, G. H. Liang, Z. Y. Mei, B. Z. Zhou, H. Zhang, J. C. Zhang, S. Chen, S. P. Zhao, Y. Tian, Z. Y. Yang, Z. C. Xiang, K. Xu, D. N. Zheng, H. Fan, Observation of critical phase transition in a generalized Aubry-Andr{\'e}-Harper model with superconducting circuit, npj Quantum Inf. \textbf{9}(1) 40 (2023).

\bibitem{shimasaki2022anomalous} T. Shimasaki, M. Prichard, H. Kondakci, J. Pagett, Y. Bai, P. Dotti, A. Cao, T. C. Lu, T. Grover and D. M. Weld, Anomalous localization and multifractality in a kicked quasicrystal, arXiv:2203.09442v2 (2022).






\bibitem{jagannathan2021fibonacci} A. Jagannathan, The Fibonacci quasicrystal: Case study of hidden dimensions and multifractality, Rev. Mod. Phys. \textbf{93}, 045001 (2021).

\bibitem{liu2022anomalous} T. Liu, X. Xia, S. Longhi and L. Sanchez-Palencia, Anomalous mobility edges in one-dimensional quasiperiodic models, SciPost Phys. \textbf{12}, 027 (2022).



\bibitem{wang2022quantum} Y. Wang, L. Zhang, W. Sun, and X. Liu, Quantum phase with coexisting localized, extended, and critical zones, Phys. Rev. B \textbf{106}, L140203 (2022).

\bibitem{lin2022general} X. Lin, X. Chen, G.-C Guo and M. Gong, General approach to tunable critical phases with two coupled chains, arXiv:2209.03060 (2022).

\bibitem{li2023emergent} S.-Z. Li, and Z. Li, Emergent Recurrent Extension Phase Transition in a Quasiperiodic Chain, arXiv:2304.11811v1 (2023).



\bibitem{roati2008anderson} G. Roati, C. D'Errico, L. Fallani, M. Fattori, C. Fort, M. Zaccanti, G. Modugno, M. Modugno and M. Inguscio, Anderson localization of a non-interacting Bose-Einstein condensate, Nature \textbf{453}, 895 (2008).




\bibitem{an2021interactions} F. A. An, K. Padavi{\'c}, E. J. Meier, S. Hegde, S. Ganeshan, J. H. Pixley, S. Vishveshwara, and B. Gadway, Interactions and Mobility Edges: Observing the Generalized Aubry-Andr{\'e} Model, Phys. Rev. Lett. \textbf{126}, 040603 (2021).

\bibitem{lahini2009observation} Y. Lahini, R. Pugatch, F. Pozzi, M. Sorel, R. Morandotti, N. Davidson and Y. Silberberg, Observation of a
Localization Transition in Quasiperiodic Photonic Lattices, Phys. Rev. Lett. \textbf{103}, 013901 (2009).


\bibitem{kraus2012topological} Y. E. Kraus, Y. Lahini, Z. Ringel, M. Verbin and O. Zilberberg, Topological States and Adiabatic Pumping
in Quasicrystals, Phys. Rev. Lett. \textbf{109}, 106402 (2012).


\bibitem{liu2020non} Y. Liu, X.-P. Jiang, J. Cao, and S. Chen, Non-Hermitian mobility edges in one-dimensional quasicrystals with parity-time symmetry, Phys. Rev. B, \textbf{101}, 174205 (2020).

\bibitem{liu2020generalized} T. Liu, H. Guo, Y. Pu and S. Longhi, Generalized Aubry-Andr{\'e} self-duality and mobility edges in non-Hermitian quasiperiodic lattices, Phys. Rev. B, \textbf{102}, 024205 (2020).

\bibitem{liu2021localization} Y. X. Liu, Q. Zhou and S. Chen, Localization transition, spectrum structure and winding numbers for one-dimensional non-Hermitian quasicrystals, Phys. Rev. B, \textbf{104}, 024201 (2021).

\bibitem{jiang2021non} X.-P. Jiang, Y. Qiao, and J. Cao, Non-Hermitian Kitaev chain with complex periodic and quasiperiodic potentials, Chinese Physics B, \textbf{30}, 077101 (2021).


\bibitem{iyer2013many} S. Iyer, V. Oganesyan, G. Refael  and D. A. Huse, Many-body localization in a quasiperiodic system, Phys. Rev. B \textbf{87}, 134202 (2013).


\bibitem{nandkishore2015many} R. Nandkishore and D. A. Huse, Many-body localization and thermalization in quantum statistical mechanics, Annu. Rev. Condens. Matter Phys. \textbf{6}, 15 (2015).

\bibitem{deng2017many} D. L. Deng, S. Ganeshan, X. P. Li, R. Modak, S. Mukerjee and J. Pixley, Many-body localization in incommensurate models with a mobility edge, Ann. Phys. \textbf{529}, 1600399 (2017).


\bibitem{hamazaki2019non} R. Hamazaki, K. Kawabata and M. Ueda, Non-Hermitian many-body localization, Phys. Rev. Lett. \textbf{123}, 090603 (2019).

\bibitem{li2016quantum} X. Li, J. H. Pixley, D. Deng, S. Ganeshan, and S. Das Sarma, Quantum nonergodicity and fermion localization in a system with a single-particle mobility edge, Phys. Rev. B \textbf{93}, 184204 (2016).


\bibitem{sarkar2021mobility} Madhumita Sarkar, R. Ghosh, A. Sen,  and K. Sengupta, Mobility edge and multifractality in a periodically driven Aubry-Andr{\'e} model, Phys. Rev. B \textbf{103}, 184309 (2021).


\bibitem{deng2019one} X. Deng, S. Ray, S. Sinha, G. V. Shlyapnikov and L. Santos, One-Dimensional Quasicrystals with Power-Law Hopping, Phys. Rev. Lett. \textbf{123}, 025301 (2019).

\bibitem{xu2020dynamical} Z. H. Xu, H. L. Huangfu, Y. B. Zhang and S. Chen, Dynamical observation of mobility edges in one-dimensional incommensurate optical lattices, New J. Phys. \textbf{22}, 013036 (2020).


\bibitem{roy2021critical} S. Roy, S. Chattopadhyay, T. Mishra and S. Basu, Critical analysis of the reentrant localization transition in a one-dimensional dimerized quasiperiodic lattice, Phys. Rev. B \textbf{105}, 214203 (2022).

\bibitem{zhang2012quantum} Z. J. Zhang, P. Q. Tong, J. B. Gong and B. W. Li, Quantum hyperdiffusion in one-dimensional tight-binding lattices, Phys. Rev. Lett. \textbf{108}, 070603 (2012).

\bibitem{li2015many} X. Li, S. Ganeshan, J. H. Pixley, and S. D. Sarma, Many-Body Localization and Quantum Nonergodicity in a Model with a Single-Particle Mobility Edge, Phys. Rev. Lett. \textbf{115}, 186601 (2015).



\end{thebibliography}

\end{document}